\documentstyle[aasms4,12pt]{article}

\begin{document}

\title{Evolution of Stellar Collision Products in Globular Clusters -- I. Head-on Collisions.}
\author{Alison Sills}
\affil{Department of Astronomy, Yale University, P.O. Box 208101, New Haven, CT, 06520-8101, USA}
\author{James C. Lombardi\altaffilmark{1}}
\affil{Center for Radiophysics and Space Research, Cornell University, Ithaca, NY 14853, USA}
\author{Charles D. Bailyn, Pierre Demarque}
\affil{Department of Astronomy, Yale University, P.O. Box 208101, New Haven, CT, 06520-8101, USA}
\author{Frederic A. Rasio} 
\affil{Department of Physics, MIT 6-201, Cambridge, MA 02139, USA}
\author{Stuart L. Shapiro\altaffilmark{2}}
\affil{Department of Physics, University of Illinois at Urbana-Champaign, 1110 West Green Street, Urbana, IL 61801, USA}

\altaffiltext{1}{also Department of Astronomy, Cornell University}
\altaffiltext{2}{also Department of Astronomy and National Center for Supercomputing Applications, University of Illinois at Urbana-Champaign}

\begin{abstract}
We explore the evolution of collisionally merged stars in the blue
straggler region of the HR diagram. The starting models for our
stellar evolution calculations are the results of the smoothed
particle hydrodynamics (SPH) simulations of parabolic collisions
between main sequence stars performed by Lombardi, Rasio and Shapiro
(1996).  Since SPH and stellar evolution codes employ different and
often contradictory approximations, it is necessary to treat the
evolution of these products carefully.  The mixture and disparity of
the relevant timescales (hydrodynamic, thermal relaxation and nuclear
burning) and of the important physical assumptions between the codes
makes the combined analysis of the problem challenging, especially
during the initial thermal relaxation of the star.  In particular, the
treatment of convection is important, and semiconvection must be
modeled in some detail.

The products of seven head-on collisions are evolved through their
initial thermal relaxation, and then through the main sequence phase
to the base of the giant branch. Their evolutionary tracks are
presented. In contrast to the assumptions in previous work, these
collision products do not develop substantial convective regions
during their thermal relaxation, and therefore are not mixed
significantly after the collision.

\end{abstract}


\section{Introduction}

Blue stragglers are main sequence stars which are more massive than
the main sequence turnoff stars in clusters. Since massive stars
evolve more quickly than low mass stars, blue stragglers must have
been born (or reborn) on the main sequence more recently than the
majority of the stars in the cluster. Numerous formation mechanisms
have been proposed (for a recent review, see Stryker 1993). The
current observations suggest that stellar collisions could be a viable
mechanism for some blue stragglers (Bailyn 1995).

In most environments, collisions between stars are very rare
events. However, the cores of some globular clusters are dense enough
that collisions between single stars are expected to happen at
dynamically significant rates (Hills \& Day 1976). Also, binary star
systems have a much larger collisional cross-section than single stars
and can enhance the number of stellar collisions (Leonard 1989,
Sigurdsson \& Phinney 1993).

It is important to include the effects of physical collisions between
stars when studying the dynamics and overall evolution of globular
clusters.  These collisions produce stars which are more massive than
most of the stars in clusters. The relative proportions of stars of
different masses affect the rate at which various dynamical processes
occur, such as core collapse and mass segregation (Elson, Hut \&
Inagaki 1987). The exchange of energy between the cluster kinetic
energy and the collisional energy can also affect dynamical
processes. Therefore, it is important to determine how many massive
stars are created by collisions.  Stellar collisions are also
important tracers of the dynamical state of globular clusters. The
collision rates and participants are determined by global cluster
properties such as density, velocity dispersion, mass function and
mass segregation. Therefore, by studying the products of stellar
collisions, we will be able to probe the dynamics, both past and
present, of globular clusters.

To fully study the consequences of stellar collisions in globular
clusters, many different lines of investigation need to be
combined. We need to know how many collisions happen in a particular
cluster, as well as what kinds of collisions occur, between which
kinds of stars, and with what frequency. These questions can be
answered with multi-mass King-Mitchie models (eg., Davies \& Benz
1995) which give the density and velocity dispersion of the
clusters. Two- and three-body scattering experiments (eg. McMillan \&
Hut 1996, Davies, Benz \& Hills 1994) can determine the probabilities
for different kinds of collision products to be produced.

In addition to determining the global collision properties of the
cluster, we need to consider individual collisions. SPH calculations
of stellar collisions relevant to stars in globular clusters (eg. Benz
\& Hills 1987, Lombardi {\it et al.} 1995, 1996 (hereafter LRS),
Sandquist {\it et al.} 1996) can provide a detailed description of the
structure and composition of the collision product. Stellar evolution
calculations can then follow the evolution of the collision product to
determine its track in a color-magnitude diagram, whether or not it
has surface abundance anomalies, and other observable properties
(e.g. rotation rate or pulsation period).

Previous evolution studies (Bailyn \& Pinsonneault 1995, Sills {\it et
al.} 1995, Ouellette \& Pritchet 1996, Sandquist {\it et al.} 1996)
used only a portion of the SPH results to model collision products.
These groups imposed the chemical profile, but not the structure, of a
collision product on an otherwise normal stellar model, and compared
the evolution of these stars to that of fully mixed models. The
results of these studies are inconsistent.  Bailyn \& Pinsonneault
(1995) and Sills {\it et al.} 1995 demonstrated that the blue
stragglers in 47 Tuc and the brightest blue stragglers in NGC 6397
could be explained only by fully mixed models. However, the SPH
simulations show that the collision product is not thoroughly mixed by
the collision.  Therefore, the blue stragglers would have to be mixed
by some process which occurs after the collision, not during it.

Leonard \& Livio (1995) proposed such a mechanism. They suggested that
the collision products will contain a large amount of thermal energy
deposited by the collision, and consequently will swell up into
something resembling a pre-main sequence star with a large convective
envelope.  Under this assumption, blue stragglers will be largely or
fully mixed due to convection when they arrive on the main sequence
despite the fact that the collision itself does not result in a
significant amount of mixing.  This scenario is based on the
speculation that stellar collision products contain enough thermal
energy to become pre-main sequence stars.  It was not clear whether
this situation would actually occur or not.  Sandquist {\it et al.}
(1996) modeled the Leonard \& Livio scenario by adding enough energy
to their initial stellar models that the stars were on the Hayashi
track, and then began their evolutionary runs. As a result, their
stars developed large surface convective zones, and also small,
short-lived convective cores. A significant fraction of the more
massive collision products were mixed as a results of the convection.

In this paper, we present a set of evolution calculations based on
more rigorous initial conditions than used in the previous
studies. Instead of imposing the collisional chemical composition
profile on an otherwise normal stellar model, we use all the stellar
structure information from the SPH results in the starting models for
our evolution calculations. We do not add energy to our models since
we do not want to make any assumptions about where in the HR diagram
the collision models should reside. In particular, we wish to
investigate the apparent discrepancy between the negligible amount of
mixing which is caused by the collision itself, and the substantial
amount of mixing which seems necessary to explain at least some blue
stragglers in some clusters.  We concentrate on the head-on collisions
in this paper. Future work will incorporate the results from
non-head-on collisions, and we will use population synthesis
techniques to compare our theoretical predictions with observations.

In section 2 we describe the different tools used in this work. In
section 3 we discuss the details of converting an SPH collision
product to a valid evolutionary starting model, and in section 4 we
present the results of the evolutionary calculations. In section 5 we
discuss the implications of these models to the origins of blue
stragglers, and detail future work to be done.

\section{Numerical Methods}

The following section briefly describes the SPH and stellar evolution
codes used in this study. We highlight the differences in underlying
physical assumptions, which must be dealt with carefully in using the
output of the SPH code as input for stellar evolution calculations.

\subsection{The Smoothed Particle Hydrodynamics Code}

SPH is a Lagrangian method and therefore ideally suited to follow
chemical mixing during the stellar collision.  LRS used a modified
version of the code developed by Rasio (1991) specifically for the
study of stellar interactions.  Fluid elements are represented by SPH
``particles'', and associated with each particle is its position ${\bf
r}$, velocity ${\bf v}$, mass $m$, entropy $s$ and a purely numerical
``smoothing length'' $h$ specifying the local spatial resolution.
Local quantities are calculated by weighted sums over nearby
particles.  The LRS collision simulations solve the equations of
motion of a large number (typically $3 \times 10^4$) of particles,
interacting by both gravitational and hydrodynamic forces, and yield
accurate structural and chemical composition profiles for the merger
remnant.

LRS performed SPH calculations of twenty-three parabolic collisions
with various mass combinations and impact parameters.  The parent
stars were main sequence stars appropriate for globular clusters, with
chemical composition profiles based on evolutionary models evolved to
15 Gyr.  The $M=0.8 M_{\odot}$ star was modeled as an $n=3$ polytrope,
the $M=0.4 M_{\odot}$ star was modeled as an $n=1.5$ polytrope, and
the $M=0.6 M_{\odot}$ star was a composite polytrope with an $n=3$
core and an $n=1.5$ envelope. In all cases, the fluid was treated as
an ideal gas with adiabatic index $\Gamma_1 = 5/3$.  The parent models
and final results are described in detail in LRS.

SPH codes, by definition, are hydrodynamic and can deal with
3-dimensional fluids which are far from hydrostatic equilibrium.  They
follow the system over dynamical timescales, which are typically on
the order of an hour for collisions involving main sequence stars in
globular clusters.  The adopted SPH code includes the effects of
shocks, but is adiabatic and otherwise neglects all radiative and heat
transport between particles, which is not significant on dynamical
timescales.  Therefore, convective energy transport is not modeled,
but the mixing effects of convective fluid motions occurring on a
dynamical timescale are included.

\subsection{The Stellar Evolution Code}
We used the Yale Rotating Evolution Code (YREC) in its non-rotating
mode for our stellar evolution calculations.  The code solves the
equations of stellar structure in one dimension using the Henyey
technique.  A detailed description of the physics included in the code
can be found in Guenther {\it et al.} (1992).  We have since updated
the opacities to include OPAL opacities for $\log T \geq 4.00$
(Iglesias \& Rogers 1996) and Alexander low temperature opacities for
$\log T \leq 4.00$ (Alexander \& Ferguson 1994) We chose a mixing
length of 1.9 pressure scale heights, the value which gives a standard
solar model using this code.  For the head-on collisions, the
assumption of non-rotation is justified, since the initial conditions
have zero angular momentum. However, the products of grazing
collisions are rotating rapidly and consequently aspherical.  In
future work involving non-head-on collisional remnants, the structural
and mixing effects of rotation will be included.

YREC assumes that the star is in hydrostatic equilibrium, and uses a
1-dimensional treatment where the star is modeled as a series of
concentric shells. One of the main concerns is to follow energy
transport closely. The timescales for evolution codes range from
thermal ($\sim 10^6$ years) to nuclear ($\sim 10^9$ years)
timescales. All of these assumptions are drastically different from
those used by the SPH code.

\section{Preparing the Collision Models for Evolution}

The different assumptions about the important physics, the timescales
and geometry treated by SPH and stellar evolution codes create
difficulties in converting the SPH results into an acceptable starting
model for use in YREC. This section describes our approach for dealing
with each of these assumptions.  Figures 1, 2 and 3 show the structure
of three collision remnants immediately after their conversion to YREC
format.

\subsection{Spherical Symmetry}

The first step is to turn the 3-dimensional particle data from SPH
into 1-dimensional radial profiles.  The SPH particles which remained
gravitationally bound to the collision product were divided into 32-45
spherical shells. The entropy and composition of the SPH particles in
each shell were averaged to give a value for that shell. Since YREC is
more accurate with more shells, the number of shells was increased to
1500 by interpolating each quantity using a cubic spline. A visual
comparison of the resulting entropy and composition profiles with the
profiles of LRS shows that no significant spurious oscillations were
introduced by this interpolation technique.  To ensure smoothness in
the derivatives of the profiles, we interpolated off of carefully
chosen mass shells not evenly spaced in mass fraction; however, the
evolutionary tracks are virtually unaffected by the particular choice
of mass shells used.

Since YREC solves the equations of stellar evolution in one dimension,
all stars are assumed to be spherically symmetric.  This constraint
poses two problems. The first is that the head-on collision products
are not spherically symmetric in their chemical composition.  We
therefore took the average of all particles in given shells.  It can
be shown (appendix A) that this is a plausible approximation.  The
second problem caused by the requirement of spherical symmetry is for
non-head on collisions (the general case) the collision products are
rotating rapidly.  However, the products of head-on collisions are not
rotating, so we treat only spherically symmetric models in this paper.
As can be seen in figure 8 of LRS, this assumption of spherically
symmetric structure for head-on collisions is quite realistic.

\subsection{Hydrostatic Equilibrium}

Starting with the entropy and composition profiles provided by the SPH
simulations of LRS, we calculated the pressure and density profiles of
each collision product by integrating the equation of hydrostatic
balance and imposing the boundary condition $P(M)=0$.  This procedure
ensures that the initial evolutionary models are in hydrostatic
equilibrium.  The pressure and density profiles were not taken
directly from the SPH results since these profiles are not accurate
enough to guarantee hydrostatic equilibrium at the level of precision
needed for YREC.

Since, at the termination of the SPH simulation, the outermost
$2$--$4\%$ of the gravitationally bound fluid is still far from
dynamical equilibrium, we do not use the SPH entropy profile in this
region but rather use the cubic spline to extrapolate from the
interior so that entropy increases with mass fraction throughout the
star.  We tested a number of schemes for approximating the profiles in
the outer few percent and found that the tracks converged after a few
evolutionary timesteps: the stellar interior is vastly more important
in determining long-lived global characteristics.

These starting models have therefore been constructed to be in precise
hydrodynamical equilibrium, although they are typically far from
thermal equilibrium.  The initial density and pressure profiles input
to YREC differ somewhat from the final SPH profiles of LRS for two
reasons.  First, the averaging over nearest neighbors in the SPH
scheme makes the central density and pressure smaller than they should
be for hydrostatic balance (these errors in the SPH scheme vanish in
the limit of perfect spatial resolution).  Secondly, the SPH density
and pressure depend on when the collision simulation is terminated,
since they continually change as fluid falls back to the stellar
surface.  Note, however, that the entropy profile is virtually
unaffected by the hydrodynamics of the outermost region and is
therefore appropriate for determining the profiles of the collision
product in complete hydrostatic equilibrium.

\subsection{Equation of State}

We used the equation of state and the previously determined profiles
of pressure and density to calculate the temperature profile of the
collision product.  While SPH assumes a fully ionized ideal gas
equation of state, YREC uses a much more sophisticated equation of
state including radiation pressure, electron degeneracy and the
effects of partial ionization. For the stars we are considering, the
ideal gas equation of state is a very good approximation, with the
only significant correction being radiation pressure.  We used the
YREC implementation of the Saha equation to determine the number of
free electrons per nucleon in each shell for temperatures less than
$\log T=6.3$, and assumed that the gas was fully ionized for hotter
regions.

\subsection{Energy Transport}

Since the SPH code did not include energy transport, it gives no
direct information about luminosity. We used the equation of radiative
transport
\begin{equation}
\frac{dT}{dm} =\frac{-3}{64\pi^2 ac} \frac{\kappa L(m)}{r^4T^3}
\end{equation} 
to give us $L(m)$, the luminosity for each shell. Here $T$ is the
temperature and $r$ is the radius of that shell. The opacity $\kappa$
includes both radiative and conductive opacities.  Equation (1) is a
good approximation at $t=0$ since none of the initial models contain
convective regions. The unusual luminosity profiles obtained reflect
the strongly non-equilibrium distribution of internal energy caused by
the collision and constitute one of the major differences from more
straightforward approximations (see figures 1, 2 and 3). For example,
the luminosity in case G is initially negative near $m/M\approx 0.25$,
which is a direct consequence of $dT/dm>0$ in this region.

\section{Evolution of the Collision Products}

Here we discuss the results of evolving the seven head-on LRS
collision products. A summary of the collision products is given in
table 1.  We emphasize the results of three particular collisions
which demonstrate important features of the evolution calculations:
case A ($0.8 M_{\odot}$ + $0.8 M_{\odot}$ , case G ($0.8 M_{\odot}$ +
$0.4 M_{\odot}$ ) and case J ($0.6 M_{\odot}$ + $0.6 M_{\odot}$ ).
Case A was chosen to represent the most massive collision product.  We
also chose to discuss two collision products (cases G and J) with
similar masses but very different collisional histories to demonstrate
that collisional history is at least as important as final mass in
determining the evolutionary track of the collision product.

\subsection{Procedure}

After converting each of the SPH models into a starting model for
YREC, the stars were evolved with timesteps of 1x$10^2$ to 5x$10^2$
years (about $10^{-4}$ of a thermal timescale) through their initial
thermal relaxation phase. This value was chosen to ensure that the
evolutionary timestep was shorter than the local thermal timescale for
most of the star. When the star reached the main sequence, the short
timestep constraint was lifted, and timesteps were chosen
automatically to maximize the efficiency of the computing. The
evolution was stopped when the star reached the giant branch.

For comparison, we also evolved normal stars with masses similar to
the collision products. To create the starting models for these stars,
we calculated fully convective polytropes ($n=1.5$) with the same
initial composition as the collision products ($Y=0.25, Z=0.001$).  We
then evolved these models from the pre-main sequence to the main
sequence and beyond to the giant branch.

To determine the maximum effect mixing could have on the evolution, we
also evolved fully mixed collision products, with helium abundances
higher than a normal globular cluster star. The chemical composition
of the SPH model was averaged over the entire star, keeping the total
helium mass the same but spreading the hydrogen and helium evenly
throughout the star. These stars were also evolved from $n=1.5$
polytropes.

\subsection{Convection and Semiconvection}

Convection is a hydrodynamical process, since it involves the motion
of fluid elements. It is also a thermal process, since convection
transports energy.  Since the SPH code does not model energy
transport, and the evolution code does not model dynamical processes,
both codes have to treat convection as an approximation.  The
convective turnover timescale for stars like the sun is typically on
the order of a month. This timescale is between the hour-long
dynamical timescale followed by the SPH code, and the century-long
timescale which is the shortest numerically stable timestep in
YREC. Therefore, convection is the most important process which spans
all timesteps and both codes, and must be treated with particular
care.

YREC, in its original form, uses the Schwarzschild criterion for
stability,
\begin{equation}
\nabla_{rad} < \nabla_{ad}
\end{equation}
where $\nabla_{rad} = (\frac {d \ln T}{d \ln P})_{rad}$ is the
radiative temperature gradient and $\nabla_{ad}$ is the adiabatic
temperature gradient, without regard for the presence of composition
gradients.  However, a gradient in the mean molecular weight $\mu$
within the star can stabilize the fluid against convection, and so it
may be necessary to use the more general Ledoux criterion,
\begin{equation}
\nabla_{rad} < \nabla_L \equiv \nabla_{ad} + \frac{\varphi}{\delta}\nabla_{\mu}
\end{equation}
where \(\nabla_{\mu} = \frac{d \ln \mu}{d \ln P}\) is the mean
molecular weight gradient, and \(\varphi=(\frac{d \ln \rho}{d \ln
\mu})_{P,T}\) and \(\delta=-(\frac{d \ln \rho}{d \ln T})_{P,\mu}\) are
quantities which depend on the equation of state.  It is possible that
a particular shell is stable against convection according to the
Ledoux criterion but unstable according to the Schwarzschild
criterion.  This situation is called semiconvection. Fluid elements in
these regions are vibrationally unstable and oscillate slowly about
their initial position with increasing amplitude (see e.g. Kippenhahn
\& Weigert 1994)

Semiconvective regions mix on a thermal timescale until the mean
molecular weight gradient becomes small enough that the region is
stable against semiconvection.  Using only the Schwarzschild criterion
is an acceptable approximation for most evolution calculations, since
the evolution timesteps are longer than the thermal
timescale. However, this is not the case in our post-collision
scenario.  One of the most important epochs that we must calculate
accurately is the initial readjustment of the star to thermal
equilibrium. By definition, this readjustment occurs on a thermal
timescale. We must therefore take timesteps which are small compared
to the thermal time.  Semiconvection must be modeled explicitly.

We use the approach of Langer (Langer, Sugimoto \& Fricke 1983,
Langer, El Eid \& Fricke 1985) to add a full treatment of
semiconvection to YREC.  First, we determine if each shell is
radiative, convective, or semiconvective according to the Ledoux and
Schwarzschild criteria above.  The chemical composition in each
semiconvective shell is updated according to the diffusion equation
\begin{equation}
\frac{\partial X_i}{\partial t} = \left( \frac{\partial X_i}{\partial t}\right)_{nuc} + \frac{\partial}{\partial M_r}\left( (4\pi r^2 \rho)^2 D \frac{\partial X_i}{\partial M_r}\right)
\end{equation}
where $X_i$ is the abundance of the i$^{th}$ element, 
\(\left( \frac{\partial X_i}{\partial t}\right)_{nuc} \) is the change in abundance due to nuclear burning, and
$D$ is the diffusion coefficient. This coefficient is given by
\begin{equation}D=\alpha \frac{2 a c T^3}{9 \kappa \rho^2 c_p}\frac{\nabla-\nabla_{ad}}{\nabla_L-\nabla}.
\end{equation}
where \( \nabla =\frac{d \ln T}{d \ln P}\) is the actual temperature
gradient in that shell and $c_p$ is the heat capacity as calculated
from the equation of state.  Langer {\it et al.} (1985) determine the
efficiency factor $\alpha$ to be of order 0.1, based on the expected
values of $\nabla-\nabla_{ad}$ from mixing length theory, and on an
analysis of the fluid dynamics by Unno (1962).  Equating the eddy
viscosity with the diffusion coefficient also gives $\alpha =0.1$.
 
The temperature gradient also depends on the convective state of each
shell of the star. This gradient, in the semiconvective region, is
determined by solving \(\nabla=\nabla_{rad}/(1+L^{sc}/L^{rad})\) for
$\nabla$, where
\begin{equation}
L^{sc}/L^{rad}=\alpha \frac{\nabla-\nabla_{ad}}{2\nabla(\nabla_L-\nabla)} \left[ (\nabla-\nabla_{ad}) - \frac{\beta (8-3\beta)\nabla_{\mu}}{32-24\beta-3\beta^2}  \right]
\end{equation}
The two luminosities, $L^{sc}$ and $L^{rad}$ represent the amount of
luminosity carried by the two energy transport mechanisms,
semiconvection and radiation respectively. Here $\beta$ is the ratio
of gas pressure to total pressure in that shell of the star.

It can be shown that the condition for dynamical stability,
\(\frac{ds}{dr} > 0 \), is the same as the Ledoux criterion for
stability against convection for any non-rotating fluid which expands
when heated (e.g. Landau \& Lifshitz 1959). If a star has an entropy
gradient which increases outwards, it will not be convective.  This
stability condition was used by LRS to determine when it was
acceptable to stop the SPH simulations. Therefore, our collision
products are radiative or semiconvective when they begin their thermal
contraction. Our evolution calculations show that they do not develop
a convective envelope until they reach the giant branch.

\subsection{Results}
The evolutionary tracks are plotted in figures 4 and 5.
Distinguishable points are marked on each track, and the times from
the collision to that point are given in table 2.  We describe the
details of the evolution of case A in the following paragraphs.  The
underlying physics of the other collisions is similar. Other specific
cases will be discussed later in this section.

Position 0 is the position of the first model after the collision.
Since the collision product is not in thermal equilibrium, it has a
larger radius and is less concentrated than a main sequence star of
the same mass. Its core is not yet hot or dense enough for any
significant fusion to occur, and so the star cannot support itself
against gravitational collapse on a Kelvin-Helmholtz timescale.  As
the star's radius shrinks, the amount of available gravitational
energy is reduced and so the luminosity also decreases.  The star's
effective temperature increases, and the star moves toward the lower
left corner of the HR diagram.

As the amount of gravitational energy which can be released by further
contraction decreases, the contraction slows. At the same time, the
temperature of the core continues to increase so that the opacity in
the core is reduced, and energy flows more rapidly.  This causes the
luminosity and the surface temperature of the star to increase and the
track turns toward the upper left of the diagram.  At the maximum
luminosity of this loop, nuclear reactions begin and the gravitational
contraction slows. When the star's luminosity is supplied by nuclear
reactions alone, the star has reached the main sequence.  The thermal
relaxation phase of evolution of these collision products occurs on a
global thermal timescale, \( t_{th}=\frac{GM^2}{2RL} \), where $M$ is
the mass of the star, and $R$ and $L$ are the radius and luminosity at
the start of the collapse phase.

The equivalent of the zero age main sequence is at position 1, and the
turnoff, or terminal age main sequence, is position 2. Between
positions 1 and 2, the star is burning hydrogen to helium in its core
and is in both hydrostatic and thermal equilibrium.  At position 2,
the central hydrogen abundance has dropped below $X=0.001$.  Between
position 2 and 3, the star is on the subgiant branch, where the energy
source for the star is gravitational contraction of the core and a
small hydrogen burning shell.  Position 3 is the base of the giant
branch, where the track turns and begins its ascent up the giant
branch.  Case G shows a hook in its main sequence evolution since this
star has a discontinuity in the hydrogen abundance of the star created
by the collision. It also develops a very small convective core about
5000 years after the collision which lasts for its entire main
sequence lifetime.  When the convective core moves outwards during the
main sequence evolution and reaches this discontinuity, a large amount
of new hydrogen is mixed into the core, so the star turns back, away
from the subgiant branch and towards the main sequence
briefly. However, that new fuel is quickly consumed and the star must
end its main sequence life.

Case J shows an interesting kink in its track at the beginning of its
evolution. Instead of collapsing immediately, the star increases in
luminosity for about 1200 years, and then collapses. This is caused by
the large hump in luminosity at $m/M_* \sim 0.9$, which radiates
outwards and leaves the star. This hump is created when the outer few
percent of the star, which was thrown off to large distances by the
collision, rains back down on the collision product and converts its
gravitational energy to thermal energy when it hits the surface.
After that energy has been radiated away, the star can collapse and
become fainter, in the same way as the other stars. The other cases do
not show this particular feature since the initial outermost
luminosity is the maximum and so the star can only become fainter.
Despite the large initial luminosities in the outer regions of these
collision products, the structure of their outer 20\% by mass is so
rarefied that the luminosity can be carried by radiative
transport. Therefore, none of the cases we studied developed
convective envelopes during the thermal collapse phase.

Figure 6 shows the tracks from the SPH models (solid line), along with
evolutionary tracks for normal stars (dotted line) and fully mixed
stars (dashed line) with the same masses as our collision products.
The normal tracks show the expected evolution for an initially
chemically homogeneous star with a normal composition for a globular
cluster ($Z=0.001, Y=0.25$). The fully mixed tracks also show the
expected evolution for chemically homogeneous stars with a composition
given for a star with the same total helium mass as the collision
product.

The SPH models are brighter than the normal stars, and with a
significant main sequence are hotter as well.  The increased helium
content of the collision product, especially the higher mass stars,
has affected the opacities and the nuclear reaction rates in the
stellar core. A higher helium content reduces the opacity in the core,
making the star brighter.  Also, this increase in luminosity coming
from the core heats up the star.  Case A shows a star which has
depleted almost all of its hydrogen fuel, and so starts its `main
sequence' life well up the hydrogen exhaustion phase, never truly
living on the main sequence. If most of the massive blue stragglers
evolve with very short lifetimes, their birthrates must be very large
to account for their observed numbers.

The fully mixed tracks are significantly different from those of the
collision products, as expected. Since the mixed stars have more
helium in their envelopes, their tracks are brighter and hotter on the
main sequence than the collision tracks.  They also live on the main
sequence for longer since these stars have more hydrogen mixed into
their interiors.  The ages for the normal and fully mixed tracks are
given in table 2.  The thermal timescale given in table 2 is the
global thermal timescale at the zero-age main sequence, and represents
an order of magnitude estimate for the pre-main sequence lifetime of
these stars.

The effects of collisional history are demonstrated in the large
differences in the tracks of case G and case J. These two stars have
similar masses, (1.133 and 1.142 $M_{\odot}$ respectively) but their
evolution differs greatly due to their very different initial
composition and stellar structure. Case G is a head-on collision
between a turnoff mass star and a star with half this mass. Since the
lower mass star begins with a lower entropy than the higher mass star,
the lower mass star essentially sinks to the center of the collision
product, creating a near-discontinuity in the composition and
temperature profiles. These anomalous initial profiles cause the star
to live on the main sequence longer than case J, which begins its life
with a core only slightly depleted in hydrogen. Case G is hotter and
brighter than case J on the main sequence because case G has a denser
core and therefore burns its fuel more quickly. Since its luminosity
is increased while its radius has remained the same, its effective
temperature also increases.  The remnant from case J, a head-on
collision between two $0.6 M_{\odot}$ stars, behaves much more like a
normal star of the appropriate mass than the case G remnant since the
initial temperature and composition profiles in case J mimic those of
the parent stars. The increased helium content of the core of case J
reduces its lifetime on the main sequence.

The four other head-on collision products which we have evolved show
similar characteristics to the three discussed above. Cases M and U
are another pair of collision products with similar masses but
different collision histories. Case M is the product of a collision
between a $0.6 M_{\odot}$ star and a $0.4 M_{\odot}$ star, resulting
in a product with a total mass of $0.946 M_{\odot}$. Case U is the
remnant of a collision between a turnoff mass star ($0.8 M_{\odot}$)
and a very low mass star ($0.16 M_{\odot}$). Again, the collision
product with the larger mass ratio (case U) is hotter and brighter.
The differences between cases U and M are less than those between
cases G and J since the parent stars of the lower mass pair (U and M)
had created less helium in their pre-collision lifetime.

Case P is not really a blue straggler, even though it is a collision
product.  This star is the product of a collision between two $0.4
M_{\odot}$ stars, and so it has a mass slightly less than the turnoff
mass of $0.8 M_{\odot}$.  The parent stars of this merger remnant have
evolved through only a small fraction of their main sequence lifetime
and so have a large amount of their hydrogen fuel left to burn. For
these two reasons, the case P star evolves on a normal track for a
turnoff mass star, and will be hidden among the normal main sequence
stars in a globular cluster.

The track for case D lies between those for case A and case G. This is
expected since case D is the product of a collision between a $0.8
M_{\odot}$ star and a $0.6 M_{\odot}$ star. The track shows the same
sort of initial increase in luminosity as case J, a significant but
fairly short-lived main sequence, and the expected evolution through
the subgiant branch to the giant branch.

\section{Summary and Discussion}

\subsection{Comparison with Pre-Main Sequence Stars}

Leonard and Livio (1995) proposed that the products of stellar
collisions would be similar to pre-main sequence stars in their
structure and evolution. They suggested that these stars gain a large
amount of thermal energy in the collision, and so they expand.  The
swollen objects resemble pre-main sequence stars in that they are not
in thermal equilibrium, do not have sufficient temperature to start
hydrogen burning, and will therefore release gravitational energy as
they contract to thermal equilibrium and the main sequence on a
thermal timescale.

While the collision products are indeed far from of thermal
equilibrium, they have some significant differences from pre-main
sequence stars.  These differences all stem from the inhomogeneity of
parent stars of the collision products, unlike the parent gas of
normal pre-main sequence stars. Since the parent stars are evolved,
they have regions with non-zero chemical composition and specific
entropy gradients. This is especially true for the stars closest to
the turnoff, and less true for lower mass stars, which have lived only
a small fraction of their main sequence lifetimes.  Also, the specific
entropy of the fluid differs between stars of different
masses. Therefore, if a collision occurs between two evolved stars of
high mass, or between two stars of different masses, the parent gas of
the collision product will not have a homogeneous entropy content.
This inhomogeneity of the entropy of the parent gas, plus the
shock-heating which occurred during the collisions, results in an
entropy gradient in the collision product. The condition for dynamical
stability is that the entropy gradient increases outwards, and so the
collision product will continue to have fluid motions on a dynamic
timescale until this condition is met. LRS's calculations show that is
condition is satisfied quickly for most of the star, usually on the
order of a day.

\subsection{Mixing}

In this section, we will concentrate on an analysis of mixing in case
A, the collision between two turnoff mass stars.  The mixing of this
collision product is most interesting, since it has the largest
semiconvective region of the seven cases presented in this paper.  The
other six cases have chemical composition profiles which are less
steep than that of case A, and smaller luminosities in the interior of
the star. These two factors cause the stars to be more radiative than
case A.

Figure 7 shows the interior structure of case A as a function of time
from 100 years after the collision to the main sequence.  The light
grey region is semiconvective, the dark grey region is convective, and
the white region is radiative.  Very little of this star is convective
at any time prior to the main sequence. Never does any helium get
mixed into the outer regions of the star, and only the inner m/M $\sim
0.1$ even becomes largely convective. Small short-lived convective
regions do appear sporadically throughout the star, but these are not
significant in mixing most of the star. This can be seen most clearly
in figure 8, which shows helium profiles of the star at 6 times during
the initial thermal relaxation phase.

The core of the collision product is initially semiconvective because
of the interplay between the luminosity at each shell in the star and
the chemical composition profile. The initial luminosity profile of
case A has a `hump' in luminosity in the inner $\sim 25\%$ of the
star. Normally, a large luminosity would cause the region to be
convective, since convective energy transport is more efficient than
radiation. However, this star also has a strong gradient in mean
molecular weight in the same region as the luminosity excess, and so
the region is only semiconvective, not convective.  As the star begins
to settle down to thermal equilibrium, this luminosity moves outwards
in the star, and the luminosity profile begins to flatten out.  The
large luminosity region leaves the innermost part of the star first,
and so the center of the star becomes radiative before the rest of the
star.  The outer boundary of the semiconvective region moves outwards
as the increased luminosity moves outwards and causes regions of the
star at higher mass fractions to become more semiconvective. The
chemical profile does not become flat until a mass fraction of $\sim
0.5$, so the mean molecular weight gradient continues to stabilize the
star.

If the luminosity excess were to move coherently out of the star, we would
expect to see a region of the same width move from the center of the star to
the edge in one thermal time. This region would be semiconvective where the
mean molecular weight gradient is large enough to stabilize the region against
convection, and convective where the gradient could not. However, the 
luminosity hump not only moves outwards, but also dissipates and heats 
the star slightly. Therefore, the excess in luminosity above that which can be
carried by radiation gets smaller as the star evolves. This means that the
semiconvective regions shrinks as it moves outwards, and by $t=10^5$ years,
radiative energy transport alone is sufficient to carry away the remaining
energy. The star becomes completely radiative at this time.

The grey cross-hatched region near $t=10^5$ years is a region in which
small regions of a few mass shells each are alternatingly convective
and semiconvective. This is an extreme version of the effect seen
elsewhere in the star, where small convective regions are interspersed
with semiconvective regions. Semiconvection is more efficient at
mixing in regions of the star where the diffusion coefficient (Eq. 5)
is large. Therefore, the mean molecular weight gradient is reduced
more quickly in some regions than others.  A reduced gradient causes
more mixing, since a smaller $\nabla_{\mu}$ means a smaller value of
$\nabla_L$ and therefore a larger diffusion coefficient. Eventually
the gradient of mean molecular weight becomes zero and convection
begins. Therefore, regions which have more efficient semiconvection
will become convective sooner. This phenomenon is discussed in more
detail by Langer {\it et al.} (1985). In the grey cross-hatched
region, the temperature gradient is very slightly larger than the
adiabatic temperature gradient (so the region is only barely
semiconvective) and the efficient semiconvective regions have reduced
their mean molecular weight gradient very close to zero. The
distinction between radiative, convective and semiconvective in this
region is becoming unclear. Eventually, however, the luminosity excess
finally leaves the star, and the entire star becomes radiative. The
compositional effects of this merging of convective state can be seen
in figure 7.

Towards the end of the thermal collapse phase, the star develops a small 
central convective zone. Stars with masses more than about a solar mass
have convective cores when on the main sequence, and this collision product
is behaving in accordance with standard stellar evolution theory.

None of the collision products modeled here, including case A, mix
significant amounts of hydrogen into their cores or helium into their
envelopes.  The stars do not become largely convective for two
reasons. The first is the influence of the gradients in mean molecular
weight, which stabilize the fluid against convection but allow
semiconvection.  This semiconvection reduces the mean molecular weight
gradient but is too slow to mix the region completely.  The second
reason that these collision products do not become convective is that
the large luminosities which are present in the outer regions of the
star at the end of the collision escape from the star in a few
thousand years since the thermal timescale in these regions is so
small. Therefore, there is little convection occurring in these stars,
and they do not mix. 

While convection must be treated carefully, the general conclusions
presented here do not depend strongly on the inclusion and details of
semiconvection. Since the outer 50\% or more of the case A product is
radiative at any given time during the initial thermal relaxation
phase (see Figure 7), this star will never completely mix. The details
of the treatment of semiconvection will determine how much mixing
occurs in the core of this star, which in turn will determine the main
sequence lifetime of this collision product. If semiconvection is
neglected entirely, the inner $1/3$ of the star will be convective and
mix immediately. Under this treatment of convection, the collision
product will have a central helium abundance of $Y=0.70$. Since the
star has used up most of its core hydrogen, its main sequence lifetime
will still be short compared to a normal star of the same
mass. Therefore, we can safely conclude that this collision product
will not mix fully during its thermal relaxation phase, and that
little hydrogen will be mixed into the core, resulting in a short main
sequence lifetime.

\subsection{Comparison with Previous Evolutionary Calculations}

Bailyn and Pinsonneault (1995) showed that the luminosity function of
blue stragglers in the globular cluster 47 Tuc matched the predicted
luminosity function for single-star collisions. However, their
evolutionary tracks were based on the assumption that a blue straggler
created by a collision between two stars was fully mixed. Both the SPH
calculations and the evolutionary tracks presented here bring this
assumption into question.  The SPH calculations show that the
collision itself does not induce a significant amount of mixing. The
evolution calculations of head-on collision products show that
convection does not play a significant role in mixing the star during
the initial thermal relaxation phase. Therefore, the blue stragglers
in 47 Tuc must have been mixed by some process we have not considered
(for example, rotational mixing). Alternatively, the total mass of the
brightest blue stragglers could be greater than twice the turnoff
mass. This might be true if triple collisions (mediated by binary-
single star or binary-binary interactions) contribute significantly to
the blue straggler birth rate.

Two groups have recently used a portion of the SPH results from
stellar collision calculations to approximate the evolution of blue
stragglers.  Sills, Bailyn \& Demarque (1995) used the SPH chemical
profile of a collision between two turnoff mass stars and imposed that
profile on an otherwise normal stellar model. Sandquist, Bolte \&
Hernquist (1996) also imposed a chemical profile on a normal stellar
model.  They then went one step further than Sills {\it et al.}:
following the suggestion of Leonard \& Livio (1995), they arbitrarily
added energy until the star was on the Hayashi track and evolved the
star from that point.  The resulting evolutionary tracks of these two
groups are generally similar, with one major difference. The track
calculated by Sandquist {\it et al.} for the equivalent of case A has
an obvious main sequence, since the inner $\sim 0.2 M_{\odot}$ of the
star became convective and mixed hydrogen into the core. Sills {\it et
al.} began the evolution on the `main sequence', and since their
starting model had depleted its core hydrogen, it evolves immediately
to the subgiant branch.

These two tracks can be compared to the equivalent track in this
paper, that of case A. In general, all three tracks occupy a similar
place in the HR diagram. However, the track presented in this paper
shows that Sills {\it et al.} underestimated and Sandquist {\it et
al.} overestimated the amount of mixing which occurs during the
thermal relaxation phase immediately after the collision. The amount
of hydrogen which is mixed into the core of the star affects both the
length of the main sequence in the HR diagram and the time spent as a
blue straggler.

Sandquist {\it et al.} did not evolve the end-product of their SPH
collision calculations directly, as we have done. For this reason,
they did not discover that, contrary to the assumption of Leonard \&
Livio (1995), the star never ends up on the Hayashi track.  Sandquist
{\it et al.} state that they believe the evolution between the end of
the SPH simulation and the pre-main sequence is unimportant, and they
neglect all effects which occur on thermal relaxation scales or
shorter.  However, we have shown that the evolution during the initial
thermal relaxation phase is quite important in determining the
structure of the star on the main sequence.  These blue stragglers
never develop a surface convection zone, and do not spend any time on
the Hayashi track. Therefore, they do not mix significantly.

\subsection{Physical Reliability of the Collision Models}

The evolutionary tracks we have presented here depend on the
assumptions made in the SPH simulations, in the evolution calculations
and in the transformation process between the two kinds of codes. Most
of these assumptions and their effect on the evolutionary tracks have
been discussed elsewhere in this paper. There are two points, however,
which have not yet been addressed.

Given the importance of the entropy profiles of the merger remnants,
one might worry that the SPH artificial viscosity might be significant
for the subsequent stellar evolution.  However, substantial changes in
the artificial viscosity parameters only weakly affect the resulting
entropy and chemical composition profiles (see \S4.2 of LRS).  Indeed,
due to the typically low velocity dispersions in globular clusters, the
stellar collisions we consider are rather weak and shocks are not the
dominant effect.  Nevertheless, our SPH code has been well tested in
the presence of shocks (eg. Rasio \& Shapiro 1991).  We have tested its
treatment of one-dimensional shocks for Mach numbers ${\cal M} \sim$ a
few, and for the artificial viscosity parameters used in LRS: we find
that a shocked region typically has a final value of $P/\rho^{5/3}$
which is only marginally ($\sim 2\%$) larger than in accurate, high
resolution simulations done with a one-dimensional SPH code.  Although
artificial viscosity can spuriously transport angular momentum and
produce entropy in differentially rotating stars where there is shear
but no shocks, that is of no concern here since we treat only head-on
collisions.
 
One crucial assumption in the SPH simulation is the use of polytropes
as parent stars in the collision. Polytropes with indices $n=1.5$ or
$n=3$ (as used in LRS) are reasonable approximations to zero age main
sequence stars, but it is not clear that they are sufficient to model
evolved stars. In particular, the increase in central mean molecular
weight due to nuclear burning causes a significant decrease in central
entropy. Since the structure of the collision product is determined by
its entropy profile, this change in entropy is important (Sills \&
Lombardi 1997). Therefore, the tracks presented here may not have
initial conditions appropriate for collisions between significantly
evolved stars.

\subsection{Future Work}

Here we have evolved only non-rotating, head-on collision
products. However, grazing incidence collisions are more likely to
occur in globular clusters.  LRS also calculated collisions with
non-zero impact parameters, resulting in rapidly rotating collision
products. Due to centrifugal support, rotating stars have lower
central temperatures and densities than their non-rotating
counterparts.  Consequently, they will evolve more slowly and spend
more time on the main sequence.  Another crucial effect rapid rotation
will have on the evolution of a star is mixing, caused by meridional
circulation. The global timescale for meridional circulation is
(Tassoul 1978) \( \tau_{c}=\tau_{therm}/\chi \), the thermal timescale
divided by the ratio of the centrifugal to gravitational
acceleration. For the collision products calculated by LRS (see their
table 3), the values of $\chi$ are typically on the order of 20-70\%
in the equatorial plane, suggesting that a significant fraction of the
star could be mixed by rotation before it reaches the main sequence.
Our fully mixed models presented in this paper represent the most
extreme case of mixing possible, from rotation or other sources, and
so provide a limiting case of the evolution of the rotating collision
products.  The differences between a mixed star and an unmixed star in
temperature, luminosity and lifetime are significant, and so future
evolutionary models of stellar collision products will need to include
the effects of rotation. 

Although grazing incidence collision products are rapidly rotating at
the end of the SPH simulations, rapid rotation rates are not seen in
blue stragglers in globular clusters and old open clusters (Mathys
1991). Therefore, if blue stragglers are indeed created in stellar
collisions, some mechanism must spin the stars down between the
collision and their appearance on the main sequence. Recent results
have shown that classical T Tauri stars, young stars with
circumstellar disks of masses down to about 0.01 M$_{\odot}$, are all
rotating with periods of about 8.5 days (Edwards {\it et al.} 1993).
One current scenario (K\"{o}nigl 1991) suggests that magnetic coupling
locks the disk to the star, transferring angular momentum to the disk
and forcing these stars to rotate with relatively long periods. If
stellar collision products have both magnetic fields and disks created
during the encounter, they could be slowed down by the same mechanism
as pre-main sequence stars. Since low mass stars typically have
magnetic fields, their collision products should also.  The collision
also throws off a few percent of the stellar material, which can
amount to as much as 0.1 M$_{\odot}$.  If even a tenth of this
material were to remain in orbit around the new star, this disk could
slow the star's rotation enough to match the observations.  These
speculations will need to be confirmed with further calculations of
the pre-main sequence evolution of blue stragglers.

Light elements, such as lithium, beryllium and boron, are burned at
low temperatures and are therefore excellent tracers of mixing and
surface convection in stars. In particular, it is possible that if
blue stragglers were created from the collision of two stars which did
not have significant surface convection zones during their first foray
on the main sequence (such as two turnoff mass stars), lithium could
still be preserved on its surface. Since we predict that the collision
products presented in this paper do not mix and do not develop surface
convection zones, we could also predict that these products should
have observable lithium. Lithium will not be destroyed in the
collision since the collision timescale is too short to burn any
significant amount of lithium, even if the temperature gets hot
enough.  Detailed models of the light element abundances, including
the effects of rotational mixing would be needed to analyze observed
abundances quantitatively.  A second popular creation mechanism for
blue stragglers is the merger of the two stars of a binary. When a
binary system merges, it is expected (Webbink 1979) that the secondary
star essentially pours itself on the primary. This results in a star
without observable lithium, since the hotter inner regions of the
secondary, where lithium has already been destroyed, presumably end up
on the surface of the blue straggler.  If more sophisticated models
confirm that lithium is indeed present on the surfaces of collision
products, we may have a tracer to distinguish between blue stragglers
created by stellar collisions and those created by the merger of
binary stars.

\subsection{Summary}

We have shown that it is possible to use the results for SPH
calculations directly as starting models for stellar evolution
calculations with YREC. In order to deal correctly with the problems
arising from the different physical assumptions of each code, the
evolution through the initial thermal relaxation must be dealt with
carefully. Small timesteps must be taken, and semiconvection must be
treated explicitly.

We have evolved seven representative cases of head-on stellar
collisions in globular clusters and discussed three of those cases in
detail.  A collision between two turnoff mass stars forms the highest
mass blue straggler possible in this creation mechanism.  The other
two cases represent typical mass blue stragglers created in very
different kinds of collisions. A comparison of the latter two tracks
demonstrate that collisional history is very important in determining
the lifetimes and positions in the HR diagram of these stars. From the
seven cases presented here, we conclude that blue stragglers which are
created by head-on collisions are not significantly mixed either
during the collision or subsequently by convection.

The interesting scenario proposed by Leonard and Livio (1995), in
which thermal energy deposited in the collision products leads to
thorough convective mixing, is not borne out by our calculations. This
poses a problem for both the low observed rotation rates of blue
stragglers and for the blue stragglers in some clusters which have
been identified as being fully mixed stars on the basis of their
location in the HR diagram. Evolution studies including rotation may
be required to resolve this problem.

\acknowledgements We would like to thank Constantine Deliyannis, Marc
Pinsonneault and Ramesh Narayan for useful discussions. This work was
supported by NASA grants NAGW-2469 and NAG5-2867 at Yale University
and NSF grant AST 91-19475 and NASA grant NAG5-2809 at Cornell
University. CB is partially supported by a National Young Investigator
grant from NSF. FAR is supported by an Alfred P. Sloan Foundation
Fellowship. Hydrodynamic simulations were performed at the Cornell
Theory Center, which receives major funding from the NSF and IBM
Corporation, with additional support from the New York State Science
and Technology Foundation and members of the Corporate Research
Institute.

\appendix
\section{Deviations from Spherical Symmetry}

Head-on collision products have spherically symmetric density and
entropy profiles when the star has returned to hydrostatic
equilibrium. However, this does not necessarily mean that the chemical
composition is also spherically symmetric (cf. fig. 10 of LRS). Some
of the SPH particles have been shock heated during the collision, and
so their entropy has been increased. The particles will settle into
their stable configuration, with entropy increasing outward. At the
end of the SPH calculation, shock heated particles will be at the same
radius as other particles which were not shock heated, but simply
started out with higher entropy. The chemical composition of these two
particles is most likely not the same, and we have a spherical shell
of constant entropy made up of particles with different chemical
composition.  Since YREC insists on spherical symmetry in all
quantities, we assigned the chemical composition of each shell to be
the average of the compositions of the particles in that shell. This
approximation is reasonable, as can be shown by considering the
stability of these different particles.

Consider a fluid element which is in mechanical equilibrium with its
surroundings, but with a different mean molecular weight.  The fluid
element must then have a different temperature than its
surroundings. In this case the differential ideal gas equation of
state can be written
\[\frac{d\mu}{\mu}=\frac{dT}{T}\] 
since \(dP\) and \(d\rho\) are equal to zero. Therefore, the element
must transfer heat to or from its surroundings. Pressure balance
occurs essentially instantaneously, and the mean molecular weight of
the element is unchanged, so the density must change to compensate for
the change in temperature of the element.  The element will now start
to sink or rise, depending on the sign of the difference in density
between the element and the surroundings. The element will continue to
sink (for example) until it reaches a layer of the star with the same
density. There it would stop, except that the mean molecular weight of
this new layer is probably not that of the fluid element. If the
chemical composition is different, the temperature must be different,
and the process continues. Eventually, the fluid element will end up
in a region with the same mean molecular weight and the slow movements
will stop. The timescale over which this process operates is the local
thermal timescale.  The details of this process will only be important
during the initial thermal relaxation phase. The effect of these
oscillations will be an averaging of chemical composition as elements
with the same composition move towards each other. Therefore, we feel
that averaging over spherical shells is a reasonable approximation,
and will not greatly affect the calculated amount of mixing or the
evolution of the star.

\clearpage 

\figcaption[Astructure.eps]{Profiles of luminosity $L$, helium
abundance $Y$, temperature $T$, radius $R$, pressure $P$ and density
$\rho$ versus mass fraction for the collision product of case A. This
collision is a head-on collision between two $0.8 M_{\odot}$ stars,
resulting in a product with a total mass of $1.498 M_{\odot}$. The
large increase in luminosity in the outer 20\% of the star is caused
by the released gravitational energy of gas which was ejected from the
star by the collision and then fell back onto the collision product.}

\figcaption[Gstructure.eps]{Same as figure 1 for case G. This
collision is a head-on collision between a $0.8 M_{\odot}$ star and a
$0.4 M_{\odot}$ star, resulting in a product with a total mass of
$1.133 M_{\odot}$. Notice the discontinuity in helium abundance at a
mass fraction of 0.2. Interior to this point lies the majority of the
smaller parent star, which has lower entropy and so sank to the center
of the collision product. The evolved core of the turnoff-mass star
has been displaced by this low-entropy material and so the maximum
helium abundance occurs away from the stellar core.}

\figcaption[Jstructure.eps]{Same as figure 1 for case J. This
collision is a head-on collision between two $0.6 M_{\odot}$ stars,
resulting in a product with a total mass of $1.142 M_{\odot}$. }

\figcaption[AGJtracks.eps]{Evolutionary tracks for three head-on
collisions: cases A, G and J.  Each track is labeled near the main
sequence.  Important evolutionary stages are marked and discussed in
the text.}

\figcaption[othertracks.eps]{Evolutionary tracks for four head-on
collisions: cases D, M, U and P. Each track is labeled as in figure
4.}

\figcaption[AGJall.eps]{Evolutionary tracks for the collision product
(solid line), a star with the same mass and the composition of a
normal globular cluster star (dotted line), and a version of the
collision product which was fully mixed before evolution was started
(dashed line). The normal and fully mixed evolutionary tracks are
shown only from the zero age main sequence onwards.  The top panel
shows case A, the middle shows case G and the bottom shows case
J. Notice that case J follows the evolutionary path of a normal star
of the same mass but starts its main sequence evolution about half-way
up the main sequence. This is expected since the star has a
composition profile quite similar to the composition profile of the
parent stars, which are partially evolved main sequence stars.}

\figcaption[zones.eps]{A plot of the convective state of case A as a
function of time during the initial thermal relaxation stage of the
star. The light grey regions are semiconvective, the dark grey regions
are convective and the white regions are radiative. The cross-hatched
region is a transition region between the semiconvective and radiative
state. The majority of the star is always radiative. The inner third
of the star develops small convective zones as semiconvection begins
to level out the chemical profile. However, these convective zones
never cover a significant fraction of the star.}

\figcaption[yprofiles.eps]{Six profiles of helium abundance as a
function of mass fraction for 6 ages during the initial thermal
relaxation stage for case A. The profiles are for ages of 0, $10^2,
10^3, 10^4, 10^5$ and $10^6$ years after the collision. Note the
appearance of the fully mixed zones which correspond to the convective
zones in figure 7. Also notice that the overall shape of the profile
does not change significantly during this phase.}

\clearpage
\plotone{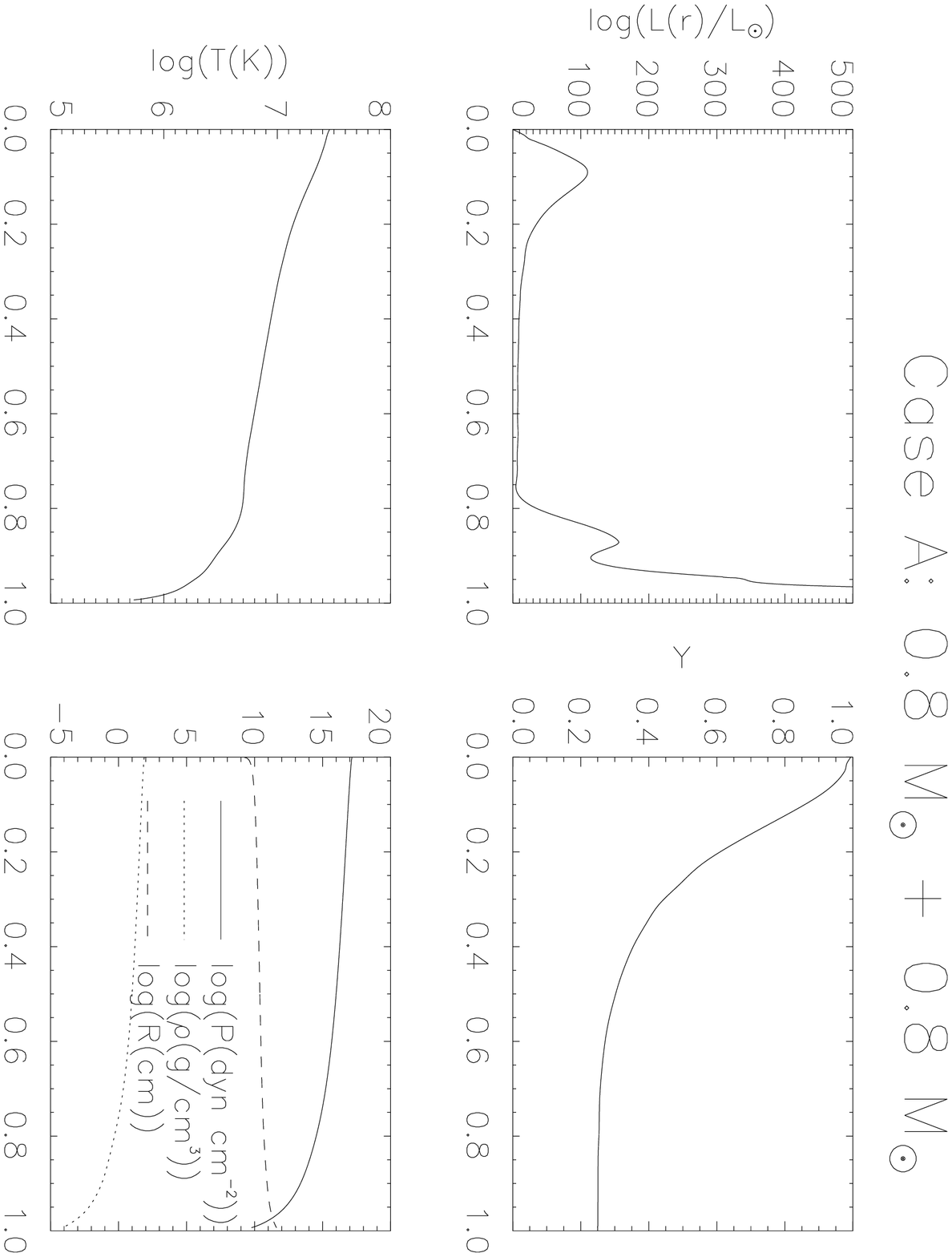}
\clearpage
\plotone{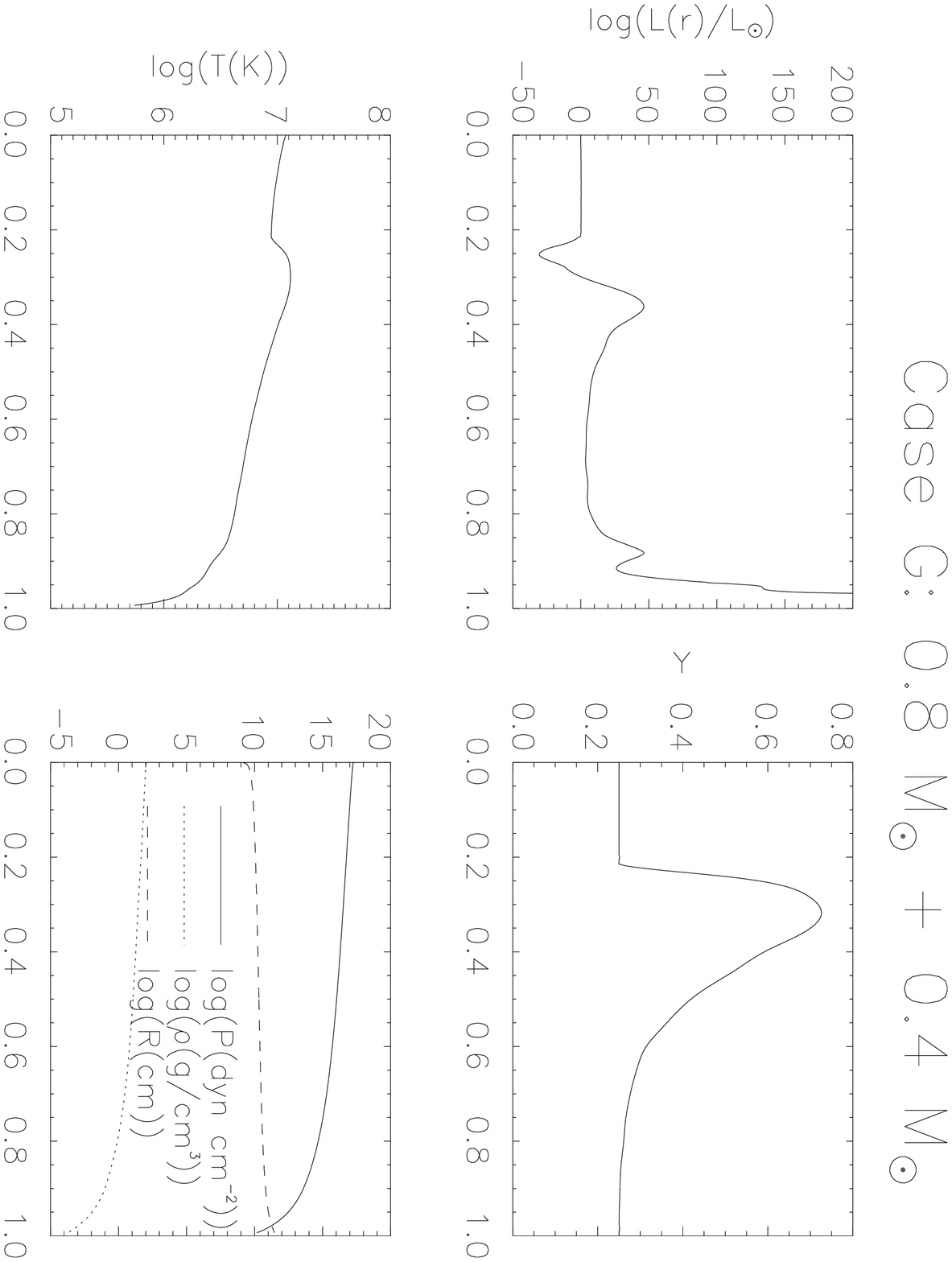}
\clearpage
\plotone{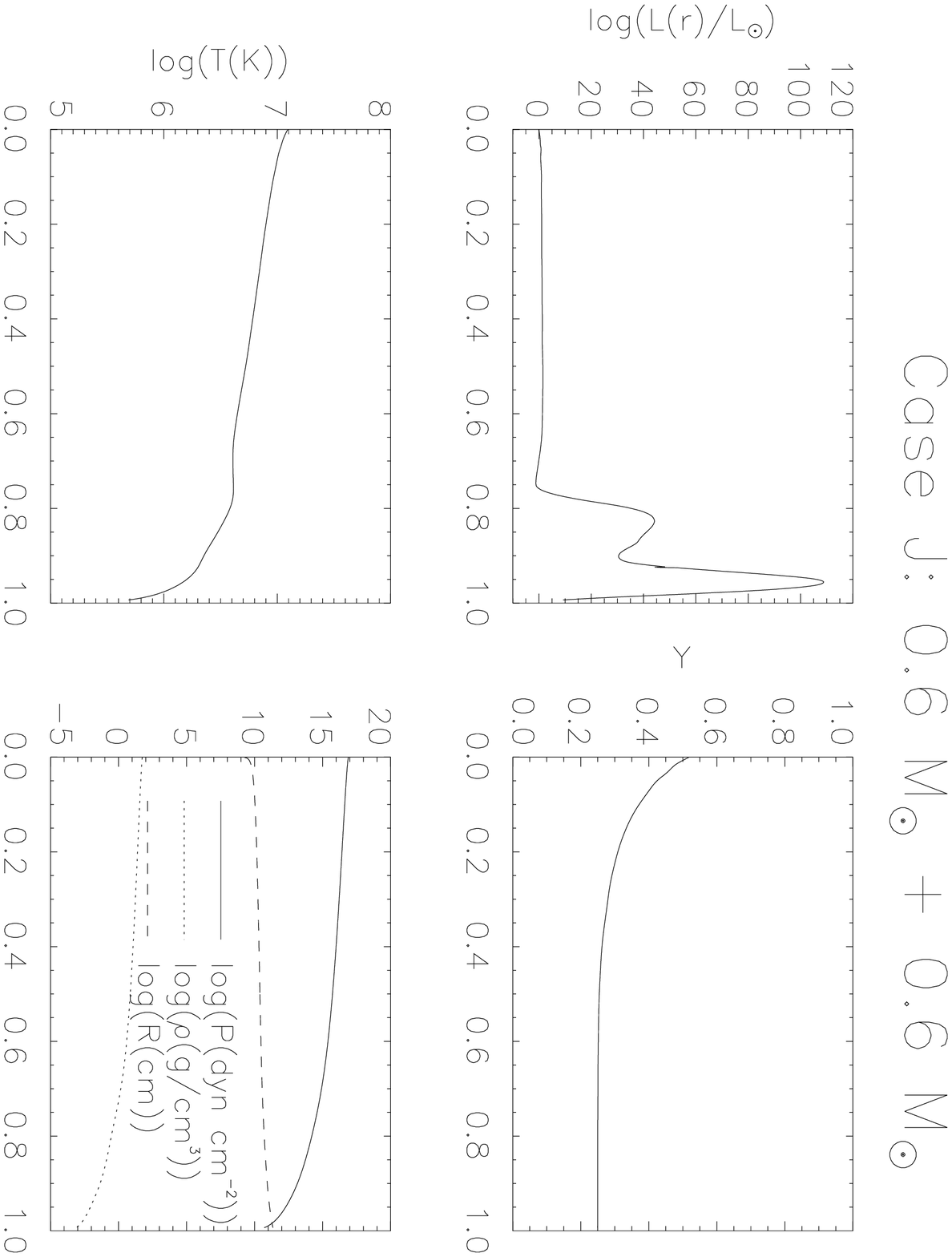}
\clearpage
\plotone{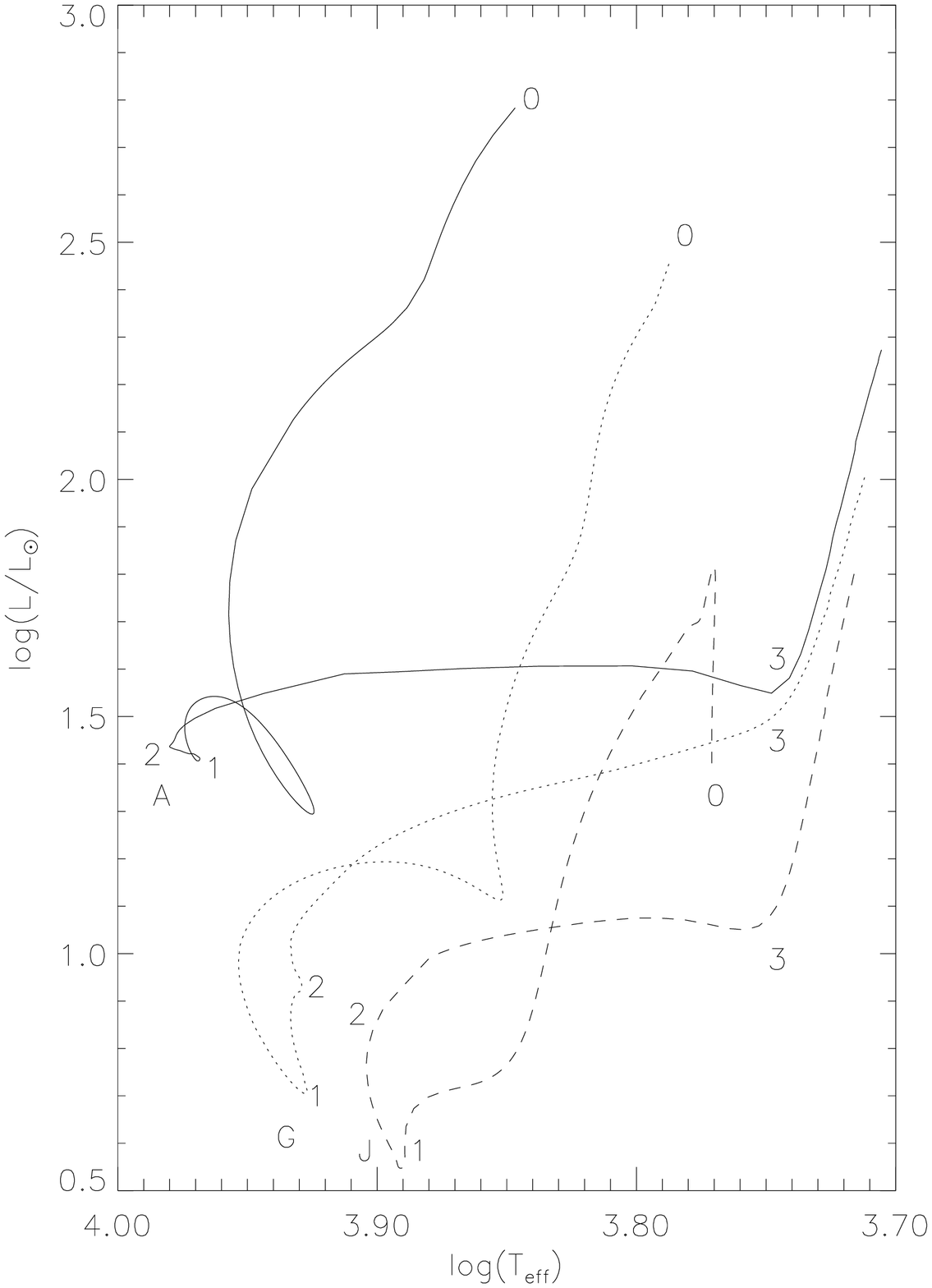}
\clearpage
\plotone{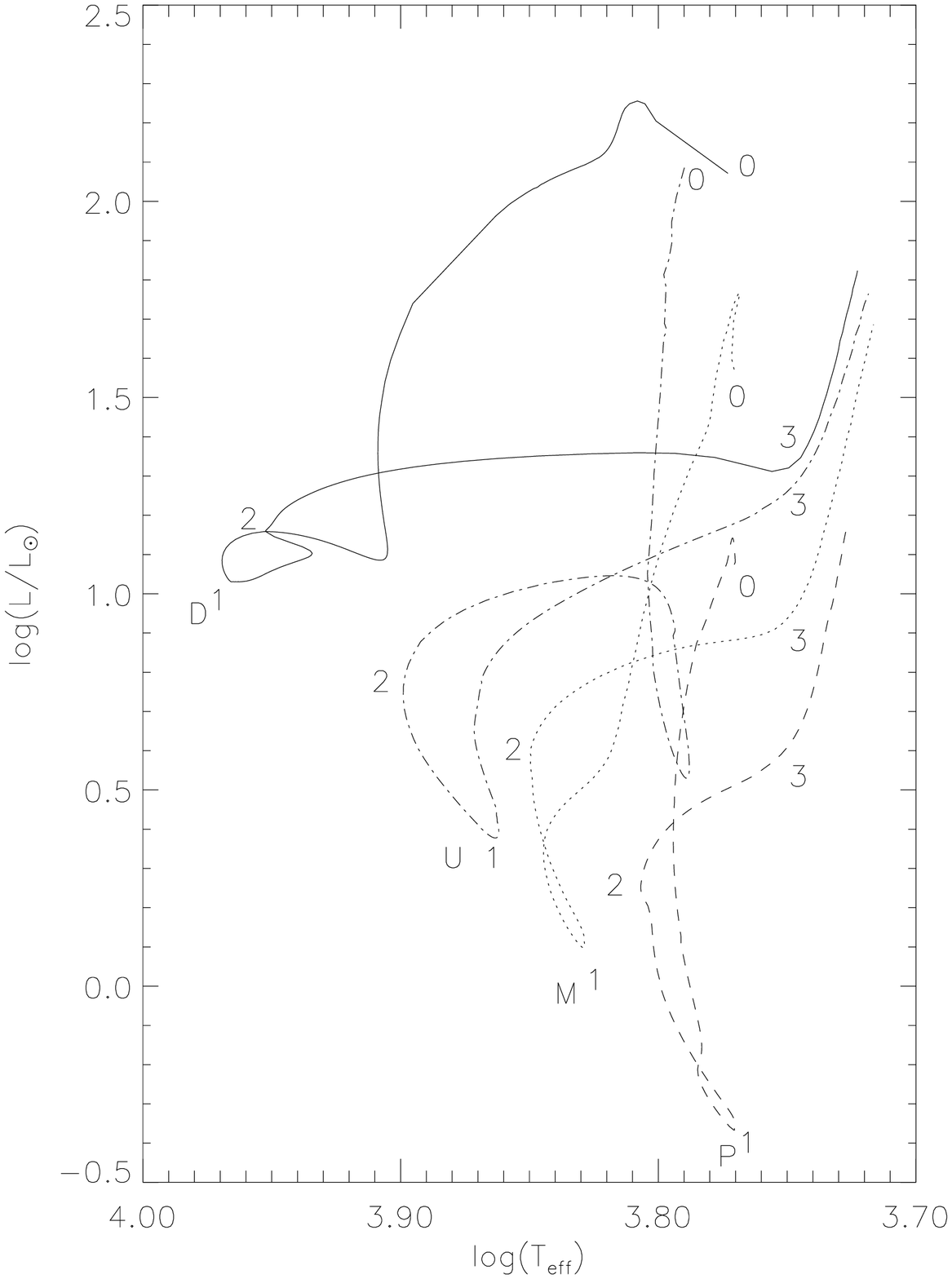}
\clearpage
\plotone{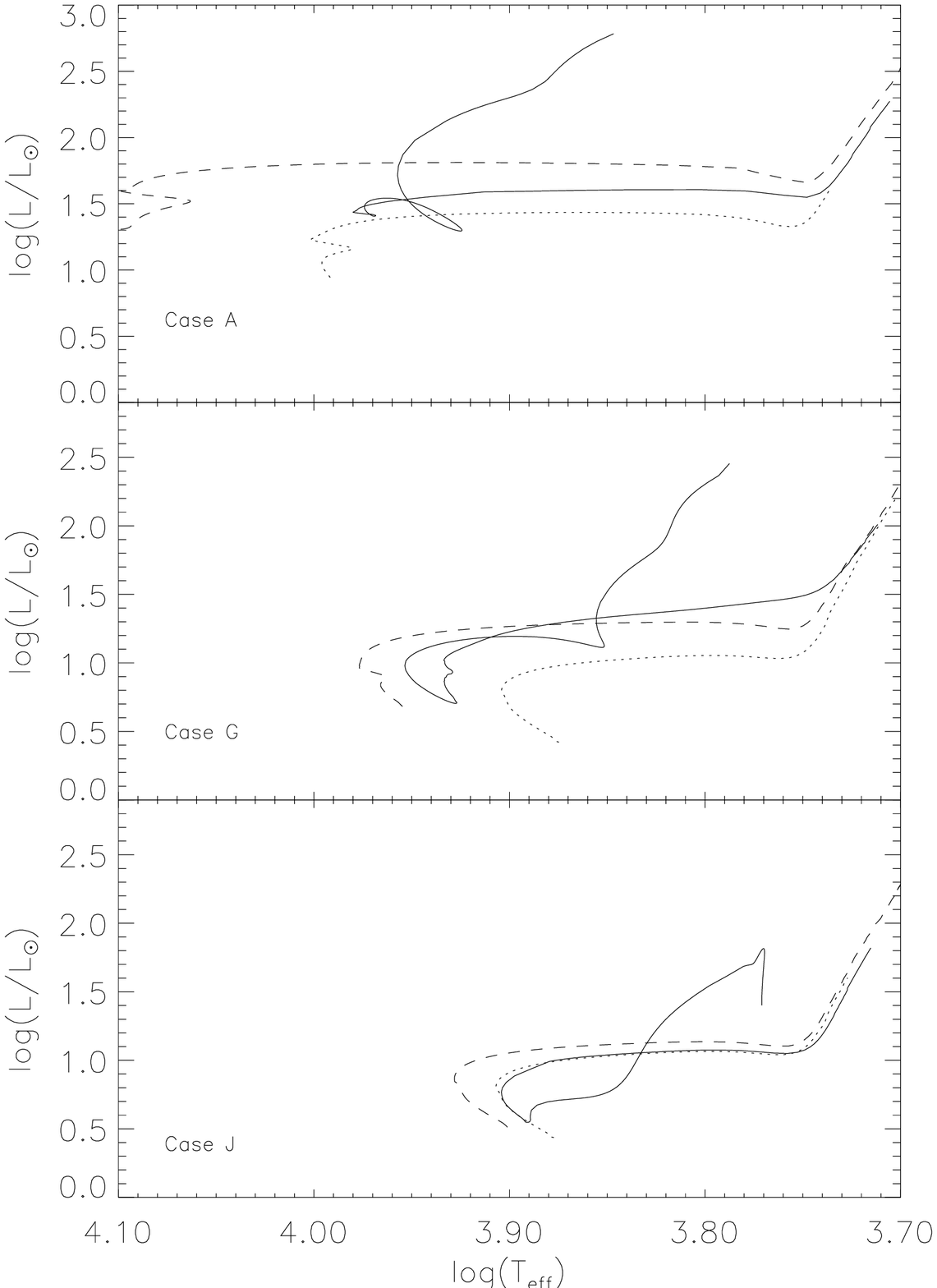}
\clearpage
\plotone{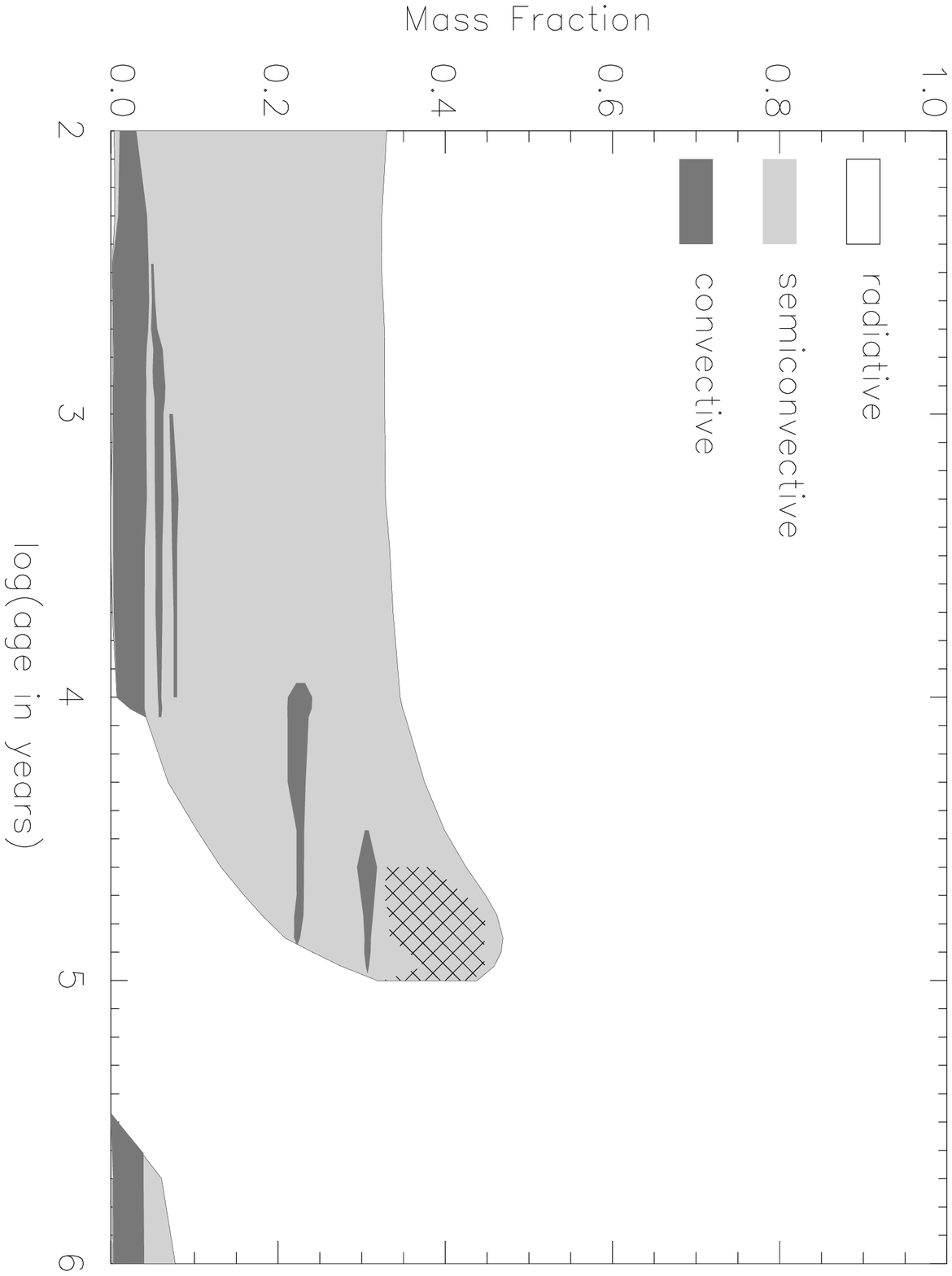}
\clearpage
\plotone{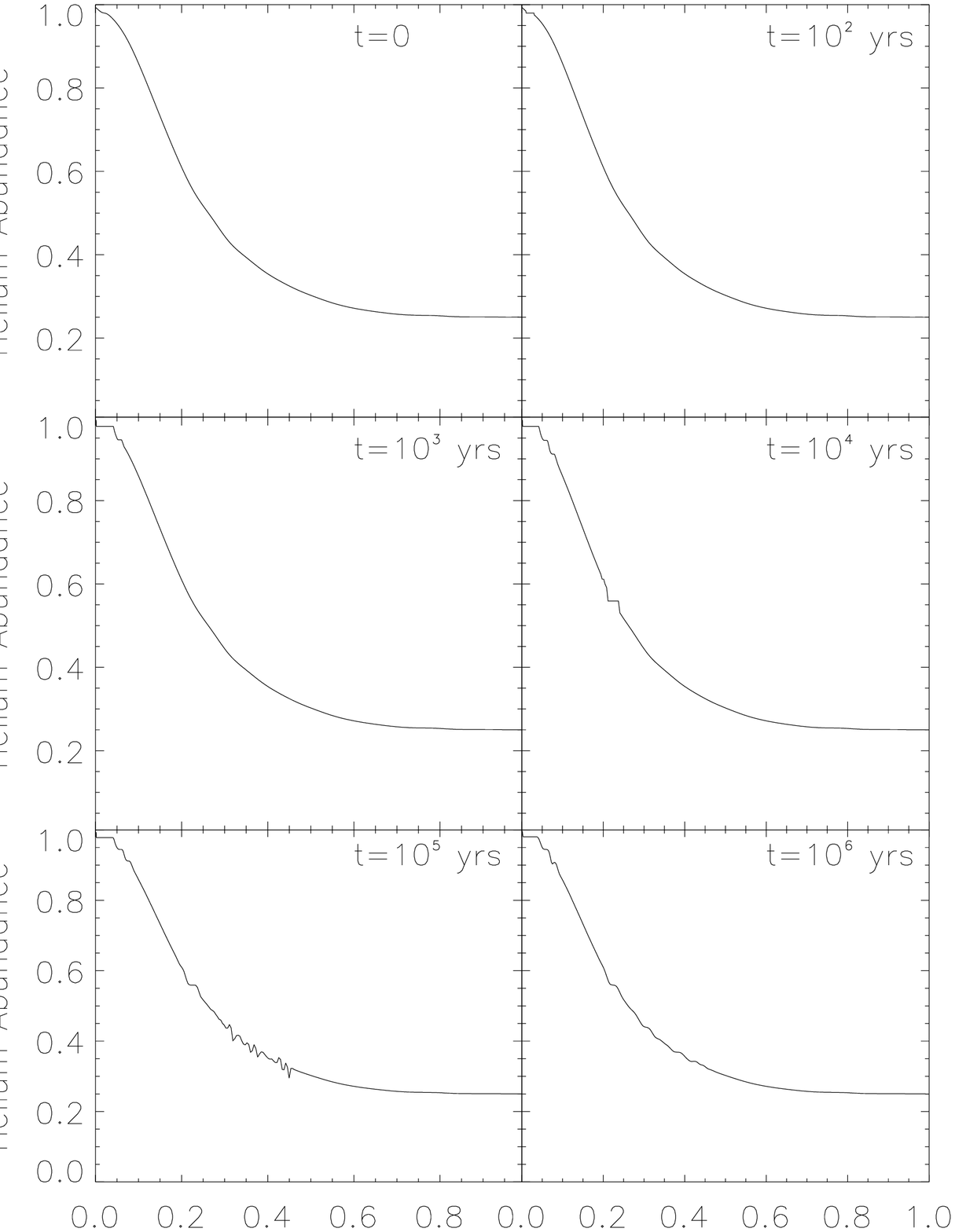}

\clearpage
\begin{deluxetable}{cccccccc}
\tablecaption{Summary of Collisions}
\tablewidth{18cm}
\tablehead{
\colhead{Case} & \colhead{M$_1$} & \colhead{M$_2$} 
& \colhead{M$_{\rm total}$} & \colhead{Helium} & \colhead{Global} & \colhead{Local Thermal} & \colhead{Local Thermal} \nl
\colhead {} & \colhead {(M$_{\odot}$)} & \colhead{(M$_{\odot}$)} & \colhead{(M$_{\odot}$)} & \colhead {Fraction} & \colhead {Thermal} & \colhead {Timescale} & \colhead {Timescale} \nl
\colhead{} &\colhead{} &\colhead{} &\colhead{} &\colhead{} &\colhead{Timescale (yrs)} &\colhead{center (yrs)} &\colhead{$m/M=0.95$ (yrs)} 
}
\startdata
A & 0.8 & 0.8 & 1.498 & 0.422 & $7.0\times 10^5$ & $7.7\times 10^6$ & $2.0\times 10^3$ \nl
D & 0.8 & 0.6 & 1.320 & 0.362 & $1.2\times 10^6$ &$ 1.1\times 10^9$ & $4.5\times 10^3$ \nl
G & 0.8 & 0.4 & 1.133 & 0.363 & $8.5\times 10^5$ & $1.7\times 10^9$ & $6.8\times 10^3$ \nl
J & 0.6 & 0.6 & 1.142 & 0.284 & $6.8\times 10^5$ & $2.1\times 10^8$ & $4.9\times 10^3$ \nl
M & 0.6 & 0.4 & 0.946 & 0.270 & $3.3\times 10^5$ & $2.8\times 10^{10}$ & $3.2\times 10^4$ \nl
P & 0.4 & 0.4 & 0.770 & 0.250 & $2.2\times 10^6$ & $2.2\times 10^{10}$ & $2.3\times 10^4$ \nl
U & 0.8 & 0.16& 0.935 & 0.385 & $2.5\times 10^4$ & $3.0\times 10^{10}$ & $2.0\times 10^4$ \nl
\enddata
\end{deluxetable}

\begin{deluxetable}{cccc}
\tablecaption{Age of Star (in years)  at Positions in HR Diagram}
\tablewidth{15cm}
\tablehead{
\colhead{} & \colhead{(1)} & \colhead{(2)} & \colhead{(3)} \nl
\colhead{Case} & \colhead{Zero Age} 
& \colhead{Terminal Age} & \colhead{Giant Branch} \nl
\colhead{}& \colhead {Main Sequence} & \colhead {Main Sequence} & \colhead {} 
}
\startdata
 A collision  & $1.12\times 10^6$ & $4.47\times 10^6$   & $8.85\times 10^7$  \nl
 D collision  & $3.51\times 10^6$ & $3.78\times 10^8$   & $6.19\times 10^8$  \nl
 G collision  & $2.82\times 10^6$ & $1.42\times 10^9$   & $1.99\times 10^9$  \nl
 J collision  & $1.41\times 10^7$ & $1.80\times 10^9$   & $2.95\times 10^9$  \nl
 M collision  & $7.11\times 10^6$ & $5.90\times 10^9$   & $7.00\times 10^9$  \nl
 P collision  & $1.42\times 10^7$ & $1.60\times 10^{10}$  & $1.75\times 10^{10}$ \nl
 U collision  & $4.31\times 10^6$ & $1.84\times 10^9$   & $3.06\times 10^9$  \nl
\hline
  &  Thermal timescale & & \nl
\hline
 A normal     & $3.86\times 10^6$ & $1.24\times 10^9$  & $1.58\times 10^9$ \nl
 G normal     & $7.85\times 10^6$ & $3.38\times 10^9$  & $4.13\times 10^9$ \nl
 J normal     & $7.63\times 10^6$ & $3.28\times 10^9$  & $4.02\times 10^9$ \nl
\hline
 A fully mixed & $1.80\times 10^6$ & $4.83\times 10^8$  & $5.62\times 10^8$ \nl
 G fully mixed & $4.60\times 10^6$ & $1.52\times 10^9$  & $2.02\times 10^9$ \nl
 J fully mixed & $6.57\times 10^6$ & $2.62\times 10^9$  & $3.26\times 10^9$ \nl
\enddata
\end{deluxetable}


\begin{references}
\reference{} Alexander, D. R. \& Ferguson, J. W. 1994, \apj, 437, 879
\reference{} Bailyn, C. D. 1995, \araa, 33, 133
\reference{} Davies, M. B., \& Benz, W. 1995, \mnras, 276, 876
\reference{} Davies, M. B., Benz, W., Hills, J. G. 1994, \apj, 424, 870
\reference{} Edwards, S., Strom, S. E., Hartigan, P., Strom, K. M., Hillenbrand, L. A., Herbst, W., Attridge, J., Merril, K. M., Probst, R., Gatley, I. 1993, \aj, 106, 372
\reference{} Elson, E., Hut, P., \& Inagaki, S. 1987, \araa, 25, 565
\reference{} Green, E. M., Demarque, P., \& King, C. R. 1987 Revised Yale Isochrones, Yale University Observatory
\reference{yrec92} Guenther, D. B., Demarque, P., Kim, Y.-C., \& Pinsonneault, M. H., 1992 \apj, 387, 372
\reference{} Hills, J. G. \& Day, C. A. 1976, {\it Astrophysical Letters}, 17, 87 
\reference{opal}Iglesias, C. A. \& Rogers, F. J. 1996, \apj, 464, 943
\reference{} K\"{o}nigl, A. 1991, \apj, 370, L39
\reference{} Kippenhahn, R. \& Weigert, A. 1994, Stellar Structure and Evolution (Berlin:Springer Verlag)
\reference{} Landau, L. D., \& Lifshitz, E. M. 1959, Fluid Mechanics (New York: Pergamon Press)
\reference{} Langer, N., El Eid, M. F., \& Fricke, K. J. 1985, \aap, 143, 179
\reference{} Langer, N., Fricke, K. J. \& Sugimoto, D. 1983, \aap, 126, 207 
\reference{} Leonard, P. J. T. 1989, \aj, 98, 217
\reference{} Leonard, P. J. T. \& Livio, M. 1995, \apjl, 447, L121
\reference{} Lombardi, J. C., Rasio, F. A., \& Shapiro, S. L. 1995, \apjl, 445, L117
\reference{} Lombardi, J. C., Rasio, F. A., \& Shapiro, S.L. 1996, \apj, 468, 797 (LRS)
\reference{} Mathys, G. 1991, \aap, 245, 467
\reference{} McMillan, S. L. W., \& Hut, P. 1996, \apj, 467, 348
\reference{} Ouellette, J. \& Pritchet, C. 1996, in ASP Conference Series \# 90 The Origins, Evolution and Destinies of Binary Stars in Globular Clusters, ed. E. F. Milone, 356
\reference{} Rasio, F. A. 1991, PhD Thesis, Cornell University
\reference{} Rasio, F. A., \& Shapiro, S. L. 1991, ApJ, 377, 559
\reference{} Sandquist, E., Bolte, M., \& Hernquist, L. 1997, \apj, 477, 335
\reference{} Sigurdsson, S., Phinney, E. S. 1993, \apj, 415, 631
\reference{} Sills, A. P., Bailyn, C. D., \& Demarque, P. 1995, \apjl, 455, L163
\reference{} Sills, A., \& Lombardi, J. C. 1997 \apj, submitted.
\reference{} Stryker, L. 1993, \pasp, 105, 1080
\reference{} Tassoul, J.-L. 1978, Theory of Rotating Stars, (Princeton:Princeton Univ. Press)
\reference{} Webbink, R. F. 1979, \apj, 227, 178 
\reference{} Unno, W. 1962, \pasj, 31, 276
\end{references}
\end{document}